\documentclass[reprint,aps,prd,nofootinbib,superscriptaddress]{revtex4-1}
\usepackage{amsmath,amssymb}
\usepackage{graphicx}
\usepackage{booktabs}

\graphicspath{{figures/}}

\begin{document}

\title{Hawking Emission from Black Holes Evaporating toward Wormholes and the Accuracy of the WKB Approximation}

\author{Bekir Can L{\"u}tf{\"u}o{\u{g}}lu}
\email{bekir.lutfuoglu@uhk.cz}
\affiliation{Department of Physics, Faculty of Science, University of Hradec Kr{\'a}lov{\'e}, Rokitansk{\'e}ho 62/26, 500 03 Hradec Kr{\'a}lov{\'e}, Czech Republic}

\date{\today}

\begin{abstract}
We revisit Hawking radiation from two black-hole families that can approach macroscopic wormhole configurations: the Simpson--Visser black-bounce geometry and the Casadio--Fabbri--Mazzacurati braneworld geometry.  The earlier analysis of these backgrounds relied on WKB greybody factors.  Here we replace that approximation by direct numerical scattering for the photon and massless Dirac channels and then recompute the emission spectra and integrated luminosities.  The qualitative picture remains the same: as the wormhole endpoint is approached the black holes cool, the total flux is strongly suppressed, and the residual emission becomes increasingly fermion dominated.  The quantitative picture, however, changes substantially.  Close to the Schwarzschild limit the WKB estimates are reasonably accurate, but far from that limit the error can be large, and near the cold endpoint it can reach orders of magnitude.  In particular, the WKB calculation can substantially overestimate the remaining luminosity precisely in the regime where the evaporation rate is most sensitive to the low-frequency tail of the greybody factors.  In a fixed-parameter Simpson--Visser half-decay estimate, the direct-greybody luminosities increase the WKB lifetime coefficient by a factor of about 85.  These results show that reliable evaporation rates for black holes evolving toward wormhole-like endpoints require direct numerical greybody factors rather than barrier-top WKB estimates alone.
\end{abstract}

\maketitle

\section{Introduction}

Hawking emission is not determined by the horizon temperature alone.  The near-horizon state is thermal, but the particles observed at infinity must cross a curvature-induced scattering barrier.  The associated transmission probabilities, or greybody factors, reshape the Planck spectrum and set the integrated luminosity \cite{Hawking1975,Page1976,Unruh1976,DasGibbonsMathur1997}.  For geometries that are close to Schwarzschild, a WKB approximation around the barrier maximum often gives a useful estimate.  For lowest multipoles, deep tunneling tails, or near-extremal temperatures, however, a small greybody-factor error can become a large luminosity error because the Hawking integral weights precisely the low-frequency part of the spectrum.

This point is important for the black holes studied recently by Bolokhov and Konoplya in Ref.~\cite{BolokhovKonoplya2025Circumvent}.  Their paper considered black holes that can evaporate toward macroscopic wormhole configurations and computed photon and fermion Hawking emission using WKB greybody factors.  The two benchmark geometries are the Simpson--Visser black-bounce metric \cite{SimpsonVisser2019} and the Casadio--Fabbri--Mazzacurati (CFM) braneworld metric \cite{CasadioFabbriMazzacurati2002}.  Both have a deformation parameter $a$, and both possess a cold limiting surface at $a=2M$.  Thus the main physical question is not merely whether the temperature vanishes, but how accurately the barrier transmission is known in the frequency range that still contributes to the small residual flux.

The present work follows the same units, the same mass normalization and the same parameter samples as Ref.~\cite{BolokhovKonoplya2025Circumvent}, but replaces the WKB greybody factors by direct numerical scattering.  We integrate the electromagnetic and massless Dirac radial equations with ingoing boundary conditions at the horizon and extract the incoming and outgoing amplitudes at large radius.  The resulting greybody factors are then inserted into the Hawking integrals.  The comparison is deliberately direct: the WKB columns in the luminosity tables are the published values of Ref.~\cite{BolokhovKonoplya2025Circumvent}, while the new columns are obtained from the direct-integration greybody factors.

The main result is twofold.  First, the qualitative evaporation picture survives: as $a$ approaches $2M$, both backgrounds cool and the emission is rapidly suppressed, with the remaining massless flux dominated by the lowest Dirac channel.  Second, the quantitative luminosities can differ substantially from WKB estimates.  The difference is modest in the Schwarzschild and mildly deformed Simpson--Visser cases, but it becomes large for the CFM branch and for near-limiting configurations.  This is exactly the regime in which the Hawking integral probes the low-frequency tail rather than the top of a smooth potential barrier.  For the illustrative fixed-$a$ Simpson--Visser half-decay estimate, the direct luminosities make the lifetime coefficient about $85$ times larger than the WKB value obtained with the same quadrature convention.

The rest of the paper is organized as follows.  Section~II summarizes the two geometries, the temperature formulas and the benchmark parameter values.  Section~III gives the electromagnetic and Dirac test-field equations, while Section~IV describes the direct-scattering method and the Hawking normalization.  Sections~V and VI present the greybody factors and emission spectra.  Section~VII compares the integrated luminosities with the WKB values of Ref.~\cite{BolokhovKonoplya2025Circumvent}, and Section~VIII gives the fixed-$a$ half-decay estimate.  Sections~IX and X discuss the physical interpretation and limitations, and Section~XI concludes.

\section{Geometries and Units}

We use the static spherically symmetric form
\begin{equation}
\label{eq:general_metric}
 ds^2=-A(r)dt^2+\frac{dr^2}{B(r)}+C(r)d\Omega_2^2 .
\end{equation}
The tortoise coordinate is defined by
\begin{equation}
\label{eq:tortoise_general}
 \frac{dr_*}{dr}=\frac{1}{\sqrt{A(r)B(r)}} .
\end{equation}
The calculations below use the same dimensionless convention as Ref.~\cite{BolokhovKonoplya2025Circumvent}: $G=c=\hbar=k_B=1$ and $M=1$.  Thus $a$ is reported as $a/M$, frequencies as $M\omega$, temperatures as $M T_H$, and luminosities as $M^2P$.

\subsection{Simpson--Visser black bounce}

The Simpson--Visser line element is \cite{SimpsonVisser2019}
\begin{equation}
\label{eq:sv_metric}
\begin{aligned}
 ds^2={}&-\left(1-\frac{2M}{\sqrt{r^2+a^2}}\right)dt^2
 +\frac{dr^2}{1-2M/\sqrt{r^2+a^2}}\\
 &+(r^2+a^2)d\Omega_2^2 .
\end{aligned}
\end{equation}
For $0\le a<2M$ it describes a black hole with outer horizon
\begin{equation}
\label{eq:sv_horizon}
 r_h=\sqrt{4M^2-a^2} .
\end{equation}
The Hawking temperature is
\begin{equation}
\label{eq:sv_temperature}
 T_H^{\rm SV}=\frac{\sqrt{4M^2-a^2}}{16\pi M^2} .
\end{equation}
At $a=0$ the metric is Schwarzschild.  At $a=2M$ the horizon coincides with the bounce surface and the temperature vanishes.
Perturbations and classical (quasinormal) spectrum of Simpson-Visser black holes have been studied in \cite{Churilova:2019cyt}.

\subsection{CFM braneworld metric}

The CFM metric \cite{CasadioFabbriMazzacurati2002} is written in the parametrization used by Ref.~\cite{BolokhovKonoplya2025Circumvent},
\begin{equation}
\label{eq:cfm_metric}
\begin{aligned}
 A(r)&=1-\frac{2M}{r},\qquad C(r)=r^2,\\
 B(r)&=\frac{\left(1-2M/r\right)\left(1-a/r\right)}{1-3M/(2r)} .
\end{aligned}
\end{equation}
The horizon remains at
\begin{equation}
\label{eq:cfm_horizon}
 r_h=2M,
\end{equation}
while the Hawking temperature is
\begin{equation}
\label{eq:cfm_temperature}
 T_H^{\rm CFM}=\frac{1}{4\pi M}\sqrt{1-\frac{a}{2M}} .
\end{equation}
The Schwarzschild geometry is recovered at $a=3M/2$, because then $B(r)=A(r)$.  The limiting surface $a=2M$ is cold.

\begin{figure*}[t]
\centering
\includegraphics[width=0.92\textwidth]{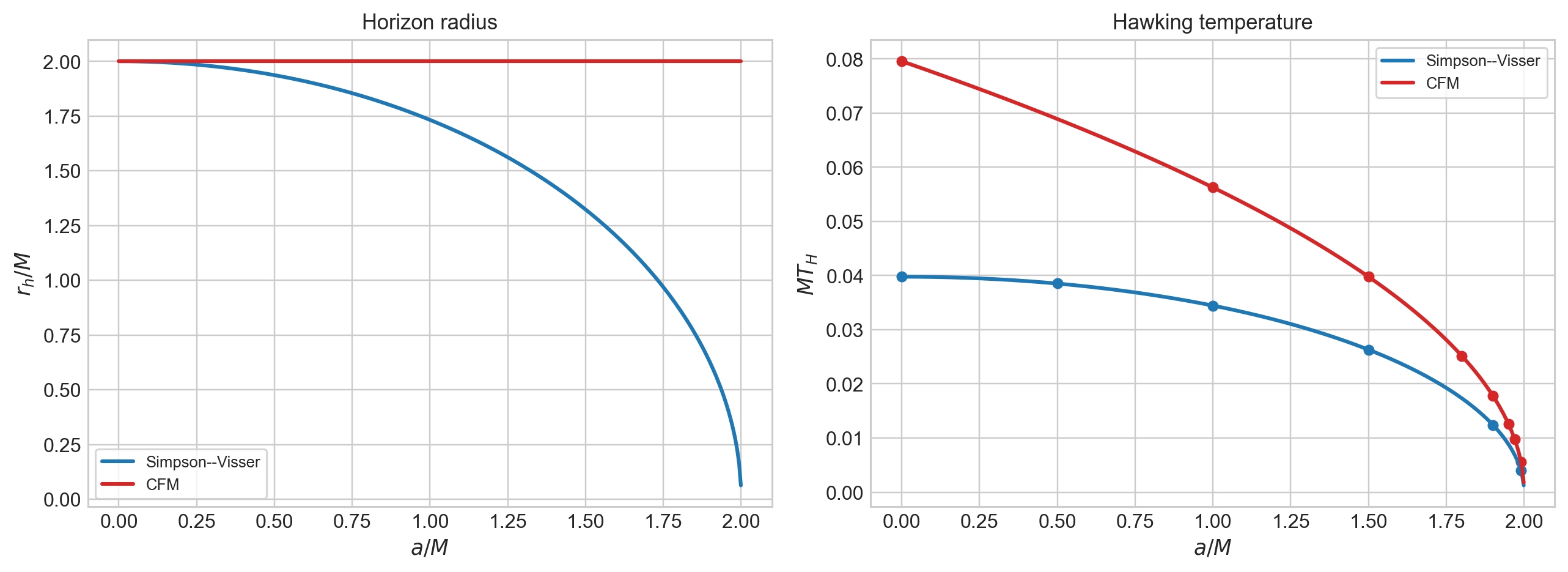}
\caption{Horizon radius and Hawking temperature for the two benchmark families.  In the Simpson--Visser case the horizon radius shrinks to zero as $a\to2M$.  In the CFM case the horizon radius stays fixed at $2M$, but the temperature still vanishes at $a=2M$.}
\label{fig:geometry}
\end{figure*}

Table~\ref{tab:geometry} lists the benchmark points used in the numerical comparison.  They are the same parameter values used in the published WKB luminosity tables of Ref.~\cite{BolokhovKonoplya2025Circumvent}.  Figure~\ref{fig:geometry} shows the corresponding horizon radii and temperatures.

Perturbations and classical spectrum of such black holes were studied in \cite{Abdalla:2006qj,Bronnikov:2019sbx,Malik:2024itg,Lutfuoglu:2026xlo}.

\begin{table}[t]
\centering
\caption{Benchmark geometries in the $M=1$ normalization.  The same parameter values are used for the direct-scattering calculation and for the comparison with the WKB luminosities of Ref.~\cite{BolokhovKonoplya2025Circumvent}.}
\label{tab:geometry}
\begingroup
\scriptsize
\begin{ruledtabular}
\begin{tabular}{cccc}
Geometry & $a/M$ & $r_h/M$ & $M T_H$ \\
Simpson--Visser & 0.00 & 2.000000 & 0.039789 \\
Simpson--Visser & 0.50 & 1.936492 & 0.038525 \\
Simpson--Visser & 1.00 & 1.732051 & 0.034458 \\
Simpson--Visser & 1.50 & 1.322876 & 0.026318 \\
Simpson--Visser & 1.90 & 0.624500 & 0.012424 \\
Simpson--Visser & 1.99 & 0.199750 & 0.003974 \\
CFM & 0.00 & 2.000000 & 0.079577 \\
CFM & 1.00 & 2.000000 & 0.056270 \\
CFM & 1.50 & 2.000000 & 0.039789 \\
CFM & 1.80 & 2.000000 & 0.025165 \\
CFM & 1.90 & 2.000000 & 0.017794 \\
CFM & 1.95 & 2.000000 & 0.012582 \\
CFM & 1.97 & 2.000000 & 0.009746 \\
CFM & 1.99 & 2.000000 & 0.005627 \\
\end{tabular}
\end{ruledtabular}
\endgroup
\end{table}

\section{Test-Field Equations}

For both backgrounds the electromagnetic and massless Dirac perturbation equations reduce to a one-dimensional scattering problem,
\begin{equation}
\label{eq:master}
 \frac{d^2\Psi}{dr_*^2}+\left[\omega^2-V(r)\right]\Psi=0 .
\end{equation}
The electromagnetic test-field potential is \cite{Konoplya:2011qq}, 
\begin{equation}
\label{eq:em_potential}
 V_\ell^{\rm EM}(r)=A(r)\frac{\ell(\ell+1)}{C(r)},
 \qquad \ell=1,2,\ldots .
\end{equation}
For a massless Dirac field we use the standard supersymmetric partner form \cite{Chandrasekhar1983,Kanti:2006ua,Cho:2003qe,Cho:2004wj,Konoplya:2007zx,Konoplya:2017tvu},
\begin{equation}
\label{eq:dirac_potential}
\begin{aligned}
 V_k^{\rm D}(r)&=W^2+\frac{dW}{dr_*},\\
 W(r)&=\frac{k\sqrt{A(r)}}{\sqrt{C(r)}},\qquad k=1,2,\ldots .
\end{aligned}
\end{equation}
The opposite chiral partner has the same transmission probability, so a single partner potential is sufficient.

The greybody factor is the transmission probability for a mode incident from infinity.  We impose a unit-amplitude ingoing wave at the horizon,
\begin{equation}
\label{eq:horizon_bc}
 \Psi\sim e^{-i\omega r_*},\qquad r\to r_h,
\end{equation}
then integrate outward.  At large radius the same solution is decomposed as
\begin{equation}
\label{eq:infinity_bc}
 \Psi\sim A_{\rm in}e^{-i\omega r_*}+A_{\rm out}e^{i\omega r_*},
 \qquad r\to\infty .
\end{equation}
With the transmitted horizon amplitude normalized to unity,
\begin{equation}
\label{eq:gamma_direct}
 \Gamma=\frac{1}{|A_{\rm in}|^2},
 \qquad
 R=\left|\frac{A_{\rm out}}{A_{\rm in}}\right|^2,
\end{equation}
so flux conservation gives $\Gamma+R=1$ as a numerical check.

\section{Numerical Method}

The direct calculation uses an adaptive Runge--Kutta integration in $r_*$, with the potential represented on a dense nonuniform radial grid.  The grid starts at $r_h+10^{-8}M$ and extends to a temperature-dependent outer radius, up to $r_{\rm max}=1050M$ for the coldest cases.  The outgoing and incoming amplitudes are extracted by averaging the plane-wave combinations
\begin{equation}
\label{eq:amplitudes}
\begin{aligned}
 A_{\rm in}&=\frac{e^{i\omega r_*}}{2}
 \left(\Psi+\frac{i}{\omega}\frac{d\Psi}{dr_*}\right),\\
 A_{\rm out}&=\frac{e^{-i\omega r_*}}{2}
 \left(\Psi-\frac{i}{\omega}\frac{d\Psi}{dr_*}\right) .
\end{aligned}
\end{equation}
across the asymptotic tail of the integration domain.  The maximum value of $|\Gamma+R-1|$ over all cached curves is $5.9\times10^{-4}$; the largest imbalances occur only in very small tail contributions, while the modes that dominate the luminosity are better conserved \cite{Tan:2026itp,Tan:2026vif,Page1976,Arbey:2025dnc}.

The frequency grid contains 120 direct-scattering points from $M\omega=2\times10^{-4}$ to $1.55$, with extra resolution below $M\omega=0.02$.  Monotone cubic interpolation is used only after the direct greybody curves are computed.  Integrated powers use a 460-point quadrature grid.  We include electromagnetic modes $\ell=1,\ldots,5$ and Dirac modes $k=1,\ldots,5$.  Higher modes are negligible for every row shown; for example, the omitted photon contribution is already smaller than the displayed digits in the cold cases and below the percent level in the hottest CFM row.

The Hawking spectra are
\begin{equation}
\label{eq:photon_spectrum}
 \frac{dE_\gamma}{dt\,d\omega}
 =\frac{1}{2\pi}\sum_{\ell=1}^{\infty}
 2(2\ell+1)\frac{\omega\Gamma_\ell^{\rm EM}(\omega)}{e^{\omega/T_H}-1},
\end{equation}
for photons, and
\begin{equation}
\label{eq:dirac_spectrum}
 \frac{dE_D}{dt\,d\omega}
 =\frac{N_D}{2\pi}\sum_{k=1}^{\infty}
 2k\frac{\omega\Gamma_k^{\rm D}(\omega)}{e^{\omega/T_H}+1}
\end{equation}
for the fermion sector.  To reproduce the normalization of Ref.~\cite{BolokhovKonoplya2025Circumvent}, we use $N_D=18$.  Equivalently, the entries in the Dirac-power tables can be divided by 18 to obtain one two-helicity massless Dirac species in the normalization used by Page. Because the black-hole temperature changes negligibly between the emission of two successive particles, the canonical ensemble provides an excellent description of the Hawking radiation process.

The masses of the Standard Model particles span a wide range, from sub-eV neutrino masses to approximately $173\,\mathrm{GeV}$ for the top quark. Nevertheless, they remain many orders of magnitude smaller than the Planck mass, $M_{\rm Pl}\simeq1.22\times10^{19}\,\mathrm{GeV}$. More importantly, the validity of the semiclassical description is controlled not by the particle masses themselves, but by the ratio of the energy carried away by each emitted Hawking quantum to the black-hole mass. As long as
\begin{equation}
\omega \ll M,
\end{equation}
where $\omega$ is the energy of the emitted particle and $M$ is the black-hole mass, the change in the black-hole parameters after a single emission is negligible. In the present case this condition is naturally fulfilled, since the Hawking temperature monotonically decreases during the evaporation process and eventually vanishes as the black hole approaches the extremal remnant. Consequently, the typical energy of the emitted quanta, $\omega\sim T_H$, also decreases and remains negligibly small compared to the black-hole mass throughout the evaporation. Therefore, the standard semiclassical Hawking formalism may be safely employed, and the backreaction of the emitted radiation on the spacetime geometry can be neglected.

The Schwarzschild check is immediate.  The Simpson--Visser row $a=0$ and the CFM row $a=1.5M$ both give
\begin{equation}
\label{eq:schwarzschild_check}
 M^2P_\gamma=3.364\times10^{-5},\qquad
 \frac{M^2P_D}{18}=4.0915\times10^{-5},
\end{equation}
in agreement with Page's direct Schwarzschild calculation at the expected accuracy \cite{Page1976}.  This check fixes both the scattering normalization and the particle-counting convention.

\section{Greybody Factors}

\begin{figure*}[t]
\centering
\includegraphics[width=0.94\textwidth]{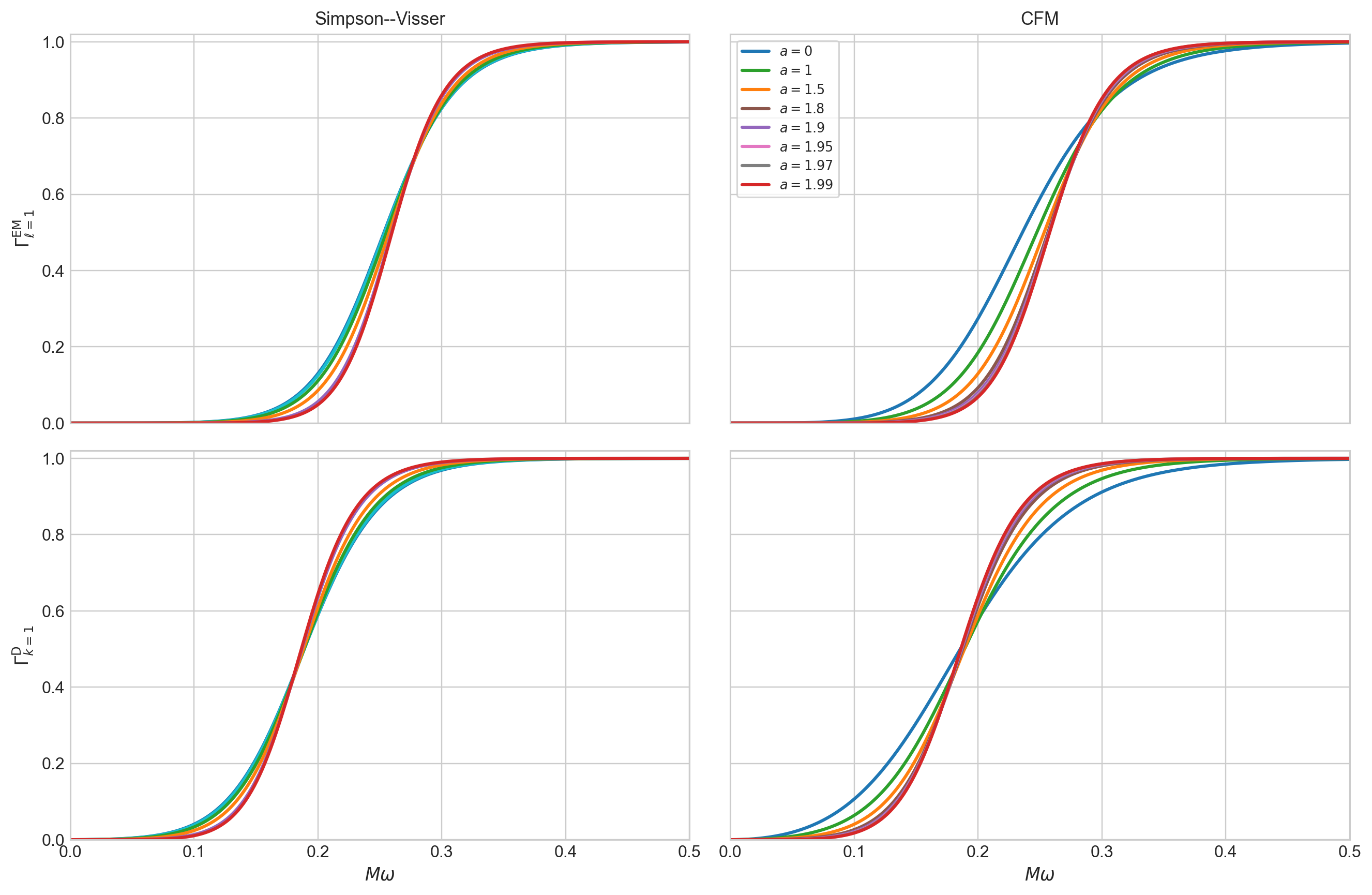}
\caption{Direct greybody factors for the dominant electromagnetic and Dirac channels, shown only up to $M\omega=0.5$ to make the threshold differences visible.  The curves are transmission probabilities obtained from numerical scattering, not WKB fits.  The dominant barriers evolve smoothly as $a$ approaches $2M$; the dramatic luminosity suppression shown below is therefore mainly thermal.}
\label{fig:greybody}
\end{figure*}

Figure~\ref{fig:greybody} shows the directly integrated greybody factors for the dominant photon and Dirac modes.  The thresholds vary only moderately across the parameter ranges.  For Simpson--Visser, the photon half-transmission frequency moves from $M\omega=0.2534$ at $a=0$ to $0.2598$ at $a=1.99M$, while the dominant Dirac threshold moves from $0.1890$ to $0.1866$.  For CFM, the photon threshold rises from $0.2358$ at $a=0$ to $0.2578$ at $a=1.99M$, while the Dirac threshold stays near $0.186$--$0.189$.  The direct thresholds are listed in Table~\ref{tab:half}.

\begin{table}[t]
\centering
\caption{Direct half-transmission frequencies for the dominant channels, defined by $\Gamma=1/2$.  All entries are in units of $M\omega$.}
\label{tab:half}
\begingroup
\scriptsize
\begin{ruledtabular}
\begin{tabular}{cccc}
Geometry & $a/M$ & EM $\ell=1$ & Dirac $k=1$ \\
Simpson--Visser & 0.00 & 0.2534 & 0.1890 \\
Simpson--Visser & 0.50 & 0.2539 & 0.1890 \\
Simpson--Visser & 1.00 & 0.2553 & 0.1889 \\
Simpson--Visser & 1.50 & 0.2575 & 0.1883 \\
Simpson--Visser & 1.90 & 0.2595 & 0.1871 \\
Simpson--Visser & 1.99 & 0.2598 & 0.1866 \\
CFM & 0.00 & 0.2358 & 0.1856 \\
CFM & 1.00 & 0.2474 & 0.1891 \\
CFM & 1.50 & 0.2534 & 0.1890 \\
CFM & 1.80 & 0.2564 & 0.1875 \\
CFM & 1.90 & 0.2572 & 0.1868 \\
CFM & 1.95 & 0.2576 & 0.1864 \\
CFM & 1.97 & 0.2577 & 0.1862 \\
CFM & 1.99 & 0.2578 & 0.1860 \\
\end{tabular}
\end{ruledtabular}
\endgroup
\end{table}

The relatively weak motion of the direct thresholds is important.  In both backgrounds the main reason for the fading flux is not the development of an opaque exterior barrier, but the fact that the thermal scale $\omega\sim T_H$ falls below the lowest barrier threshold.  Once this happens, the integrated luminosity is controlled by the low-frequency tunneling tail of the greybody factor.  This is also the region where a barrier-top WKB expansion is least reliable.

\section{Emission Spectra}

\begin{figure*}[t]
\centering
\includegraphics[width=0.94\textwidth]{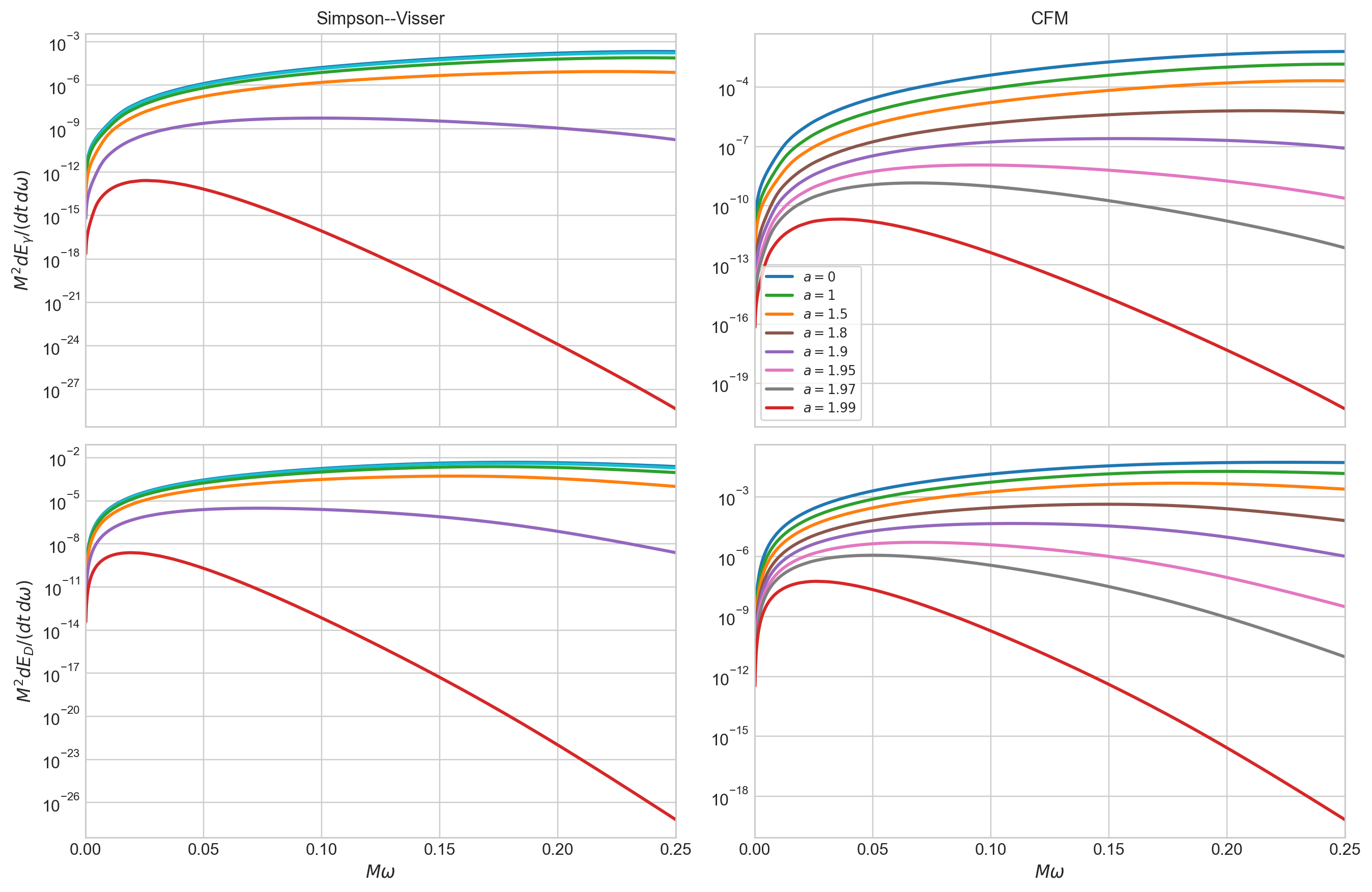}
\caption{Direct Hawking spectra for photons and for the $N_D=18$ massless Dirac normalization used for comparison with Ref.~\cite{BolokhovKonoplya2025Circumvent}.  The spectra are computed from direct greybody factors and the analytic Hawking temperatures in Eqs.~\eqref{eq:sv_temperature} and \eqref{eq:cfm_temperature}.  A common cutoff $M\omega\le0.25$ is used to focus on the emission region and to omit the high-frequency numerical tail of the coldest curves.}
\label{fig:spectra}
\end{figure*}

Figure~\ref{fig:spectra} shows the direct Hawking spectra.  The peaks move to lower frequency and lower amplitude as the limiting surface is approached.  The photon channel is especially sensitive because its lowest mode has a higher threshold and a stronger low-frequency suppression.  The massless Dirac channel remains comparatively larger near the endpoint, even though it is also absolutely quenched.

For the Simpson--Visser branch the suppression is monotonic and severe.  The photon luminosity drops from $3.36\times10^{-5}$ at $a=0$ to $6.14\times10^{-10}$ at $a=1.9M$ and $7.91\times10^{-15}$ at $a=1.99M$.  The Dirac sector drops from $7.36\times10^{-4}$ to $2.85\times10^{-7}$ and then to $6.28\times10^{-11}$ at the same points.  For the CFM branch, the high-temperature $a=0$ case is much brighter than Schwarzschild; the direct result is $M^2P_\gamma=1.81\times10^{-3}$ and $M^2P_D=1.74\times10^{-2}$.  Moving toward $a=2M$ again quenches the emission, giving $M^2P_\gamma=8.69\times10^{-13}$ and $M^2P_D=2.04\times10^{-9}$ at $a=1.99M$.

\section{Luminosity Comparison with the WKB Calculation}

As a first step, we compare the grey-body factors obtained numerically in the present work with those calculated in Ref.~\cite{BolokhovKonoplya2025Circumvent} using the higher-order WKB approximation \cite{Schutz:1985km,Konoplya:2003ii,Konoplya:2019hlu,Matyjasek:2017psv,Konoplya:2026rjh}. 

The WKB method determines the transmission probability through the effective potential barrier from its values and derivatives at the maximum.  It is based on the expansion of the potential in the maximum and matching it with the WKB series near the horizon and at spatial infinity. The transmission coefficient is written in the form
\begin{equation}
|T(\omega)|^2=\left(1+e^{2\pi K}\right)^{-1},
\end{equation}
where the quantity \(K\) is represented as an asymptotic series whose leading term depends on the height and curvature of the potential barrier and its second derivative, while higher-order corrections involve progressively higher derivatives evaluated at its peak. In the present work, we use the data obtained in \cite{BolokhovKonoplya2025Circumvent}, where the WKB expansion was carried out up to the sixth order. This choice is motivated by the fact that the sixth-order WKB approximation has frequently been found to provide the highest accuracy among the available WKB orders \cite{Konoplya:2004uk,Skvortsova:2023zmj,Konoplya:2006rv,Jawad:2023zlu,Albuquerque:2023lhm,Konoplya:2006ar,Chen:2021gwy,Lutfuoglu:2025ohb,Bolokhov:2024ixe}. 

It should be emphasized that the WKB method is an asymptotic technique whose convergence is not guaranteed order by order, and increasing the WKB order does not necessarily lead to a systematic improvement in accuracy  \cite{Konoplya:2024kih,Bolokhov:2023bwm,Fernando:2016ftj,Wongjun:2019ydo,Bolokhov:2026dzn,Lutfuoglu:2026pgn,Skvortsova:2026jtx,Konoplya:2007yy,Momennia:2018hsm,Guo:2020caw,Eniceicu:2019npi,Bolokhov:2024bke,Konoplya:2010kv,Karmakar:2023cwg,Breton:2017hwe,Skvortsova:2024eqi,Lutfuoglu:2025hjy,Bolokhov:2026kqu,Konoplya:2009hv,Skvortsova:2026unq,Skvortsova:2025cah,Konoplya:2023ahd,Bolokhov:2025egl,Konoplya:2019ppy}. Nevertheless, the method has proven remarkably successful for calculating the dominant quasinormal modes of black holes and other compact objects, where it often demonstrates excellent agreement with highly accurate numerical approaches \cite{Bolokhov:2023dxq,Konoplya:2010vz,Malik:2025erb,Bolokhov:2026uol,Lutfuoglu:2025ljm,Skvortsova:2026idf,Kodama:2009bf,Konoplya:2024hfg,Bolokhov:2025lnt,Malik:2026lfj,Bolokhov:2026dfg,Lutfuoglu:2025ljm,Skvortsova:2024wly,Konoplya:2023ppx,Skvortsova:2024atk}.

The situation is more subtle for grey-body factors. In contrast to the quasinormal-mode problem, where the convergence of the WKB series can be significantly improved through the use of Padé resummation techniques \cite{Matyjasek:2017psv,Matyjasek:2026yiu,Konoplya:2023moy}, such an enhancement is generally unavailable in scattering calculations. As a consequence, the accuracy of the WKB approximation for transmission and reflection coefficients may deteriorate considerably, particularly in regimes where the effective potential differs substantially from the simple Schwarzschild barrier.

As we shall demonstrate below, the WKB results remain reasonably close to the numerically exact grey-body factors in the Schwarzschild limit. However, the discrepancy grows progressively as the system approaches the near-extremal regime. In this region even relatively modest inaccuracies in the transmission coefficients can accumulate and lead to substantial deviations in integrated quantities, such as the Hawking emission rates. Consequently, the resulting radiation spectra may differ by several orders of magnitude from the accurate numerical predictions, highlighting the necessity of employing fully numerical methods when studying Hawking radiation in near-extremal configurations.

\begin{figure*}[t]
\centering
\includegraphics[width=0.94\textwidth]{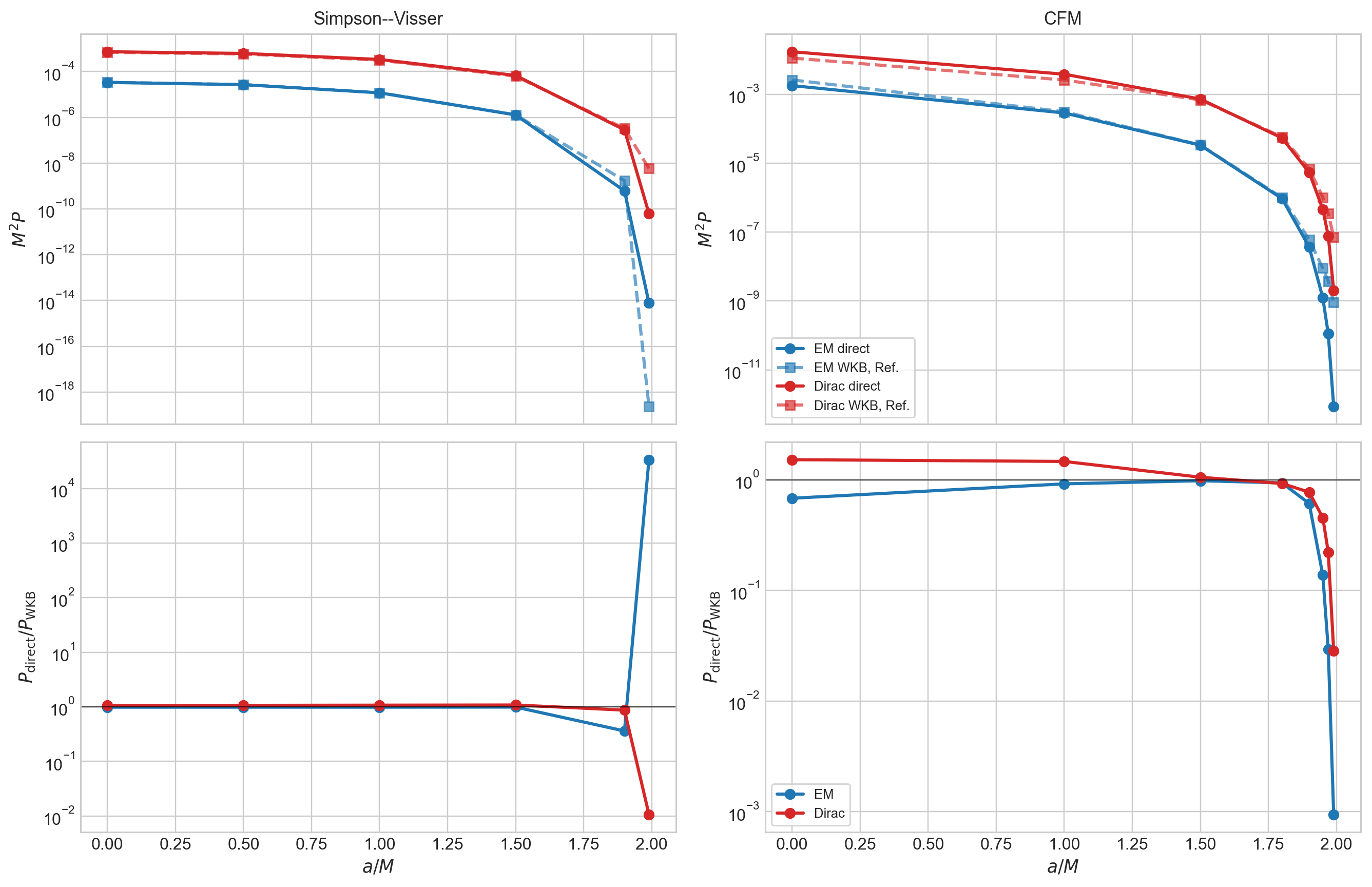}
\caption{Direct luminosities compared with the published WKB luminosities of Ref.~\cite{BolokhovKonoplya2025Circumvent}.  The upper panels show absolute powers.  The lower panels show $P_{\rm direct}/P_{\rm WKB}$.  The largest relative discrepancies occur when the thermal scale is far below the dominant barrier threshold, so that the Hawking integral is controlled by low-frequency tunneling rather than barrier-top transmission.}
\label{fig:power_comparison}
\end{figure*}

Tables~\ref{tab:sv_power} and~\ref{tab:cfm_power}, and Fig. \ref{fig:power_comparison} compare the direct luminosities with the WKB luminosities quoted in Ref.~\cite{BolokhovKonoplya2025Circumvent}.  The relative column is
\begin{equation}
\label{eq:relative_difference}
 \Delta_P=\frac{P_{\rm direct}-P_{\rm WKB}}{P_{\rm WKB}} .
\end{equation}
Thus a negative number means that the WKB calculation overestimates the direct power.

\begin{table*}[t]
\centering
\caption{Simpson--Visser integrated powers.  Direct values are obtained from numerical greybody factors.  WKB values are the published luminosities of Ref.~\cite{BolokhovKonoplya2025Circumvent}.  The Dirac normalization uses $N_D=18$, as in Eq.~\eqref{eq:dirac_spectrum}.}
\label{tab:sv_power}
\begingroup
\scriptsize
\setlength{\tabcolsep}{4pt}
\begin{ruledtabular}
\begin{tabular}{ccccccc}
$a/M$ & $P_\gamma^{\rm direct}$ & $P_\gamma^{\rm WKB}$ & $\Delta_\gamma$ & $P_D^{\rm direct}$ & $P_D^{\rm WKB}$ & $\Delta_D$ \\
0.00 & $3.3639\times10^{-5}$ & $3.4189\times10^{-5}$ & $-1.6\%$ & $7.3648\times10^{-4}$ & $6.9585\times10^{-4}$ & $5.8\%$ \\
0.50 & $2.6812\times10^{-5}$ & $2.7247\times10^{-5}$ & $-1.6\%$ & $6.2187\times10^{-4}$ & $5.8492\times10^{-4}$ & $6.3\%$ \\
1.00 & $1.1815\times10^{-5}$ & $1.2009\times10^{-5}$ & $-1.6\%$ & $3.3942\times10^{-4}$ & $3.1676\times10^{-4}$ & $7.2\%$ \\
1.50 & $1.2809\times10^{-6}$ & $1.2956\times10^{-6}$ & $-1.1\%$ & $6.7844\times10^{-5}$ & $6.2885\times10^{-5}$ & $7.9\%$ \\
1.90 & $6.1428\times10^{-10}$ & $1.7016\times10^{-9}$ & $-63.9\%$ & $2.8490\times10^{-7}$ & $3.2785\times10^{-7}$ & $-13.1\%$ \\
1.99 & $7.9078\times10^{-15}$ & $2.3534\times10^{-19}$ & $3360148.1\%$ & $6.2811\times10^{-11}$ & $5.8693\times10^{-9}$ & $-98.9\%$ \\
\end{tabular}
\end{ruledtabular}
\endgroup
\end{table*}

\begin{table*}[t]
\centering
\caption{CFM integrated powers in the same notation as Table~\ref{tab:sv_power}.  The Schwarzschild point is $a=1.5M$, not $a=0$.}
\label{tab:cfm_power}
\begingroup
\scriptsize
\setlength{\tabcolsep}{4pt}
\begin{ruledtabular}
\begin{tabular}{ccccccc}
$a/M$ & $P_\gamma^{\rm direct}$ & $P_\gamma^{\rm WKB}$ & $\Delta_\gamma$ & $P_D^{\rm direct}$ & $P_D^{\rm WKB}$ & $\Delta_D$ \\
0.00 & $1.8105\times10^{-3}$ & $2.6496\times10^{-3}$ & $-31.7\%$ & $0.01743$ & $0.01143$ & $52.5\%$ \\
1.00 & $2.9061\times10^{-4}$ & $3.1515\times10^{-4}$ & $-7.8\%$ & $3.8741\times10^{-3}$ & $2.6347\times10^{-3}$ & $47.0\%$ \\
1.50 & $3.3639\times10^{-5}$ & $3.4189\times10^{-5}$ & $-1.6\%$ & $7.3648\times10^{-4}$ & $6.9585\times10^{-4}$ & $5.8\%$ \\
1.80 & $9.4957\times10^{-7}$ & $1.0094\times10^{-6}$ & $-5.9\%$ & $5.4500\times10^{-5}$ & $5.8678\times10^{-5}$ & $-7.1\%$ \\
1.90 & $3.7188\times10^{-8}$ & $6.0733\times10^{-8}$ & $-38.8\%$ & $5.4160\times10^{-6}$ & $6.9840\times10^{-6}$ & $-22.5\%$ \\
1.95 & $1.2419\times10^{-9}$ & $9.0109\times10^{-9}$ & $-86.2\%$ & $4.6520\times10^{-7}$ & $1.0204\times10^{-6}$ & $-54.4\%$ \\
1.97 & $1.1148\times10^{-10}$ & $3.7753\times10^{-9}$ & $-97.0\%$ & $7.8019\times10^{-8}$ & $3.5051\times10^{-7}$ & $-77.7\%$ \\
1.99 & $8.6878\times10^{-13}$ & $9.1781\times10^{-10}$ & $-99.9\%$ & $2.0446\times10^{-9}$ & $7.1637\times10^{-8}$ & $-97.1\%$ \\
\end{tabular}
\end{ruledtabular}
\endgroup
\end{table*}

The Simpson--Visser comparison is benign until the very cold regime.  At $a=0$, $0.5M$ and $1.0M$, the direct photon powers are only about $1.6\%$ below the WKB values, while the direct fermion powers are about $6$--$7\%$ higher.  At $a=1.5M$ the deviations remain small: $-1.1\%$ for photons and $+7.9\%$ for fermions.  The near-limiting points are different.  At $a=1.9M$ the direct photon power is $64\%$ below the WKB value and the direct fermion power is $13\%$ below.  At $a=1.99M$ both absolute powers are extremely small; the photon relative difference is numerically huge because the WKB photon denominator is already $2.35\times10^{-19}$, while the direct value is $7.91\times10^{-15}$.  The fermion comparison is physically clearer there: direct scattering gives $6.28\times10^{-11}$, about $1\%$ of the WKB value.

The CFM differences are larger over most of the branch.  At $a=0$ the direct photon luminosity is $31.7\%$ below the WKB value, while the direct fermion luminosity is $52.5\%$ above it.  The Schwarzschild point $a=1.5M$ behaves as expected, with only a percent-level photon difference and a $5.8\%$ fermion difference.  Near the cold endpoint the WKB method overestimates both channels by increasingly large factors.  At $a=1.95M$, the direct photon power is only $13.8\%$ of the WKB value and the direct Dirac power is $45.6\%$ of the WKB value.  At $a=1.99M$, these ratios become $9.5\times10^{-4}$ and $2.85\times10^{-2}$, respectively.

Tables~\ref{tab:sv_partial} and~\ref{tab:cfm_partial} give the direct partial-mode decomposition.  They show that the discrepancies in the total powers are not caused by a missing large number of high multipoles.  In the cold rows the luminosity is almost entirely the lowest electromagnetic or Dirac mode.  In the hot CFM row $a=0$, higher modes contribute visibly, but the calculation includes them through $\ell,k=5$ and the total is already stable at the displayed precision.

\begin{table*}[t]
\centering
\caption{Direct Simpson--Visser partial powers.  The photon columns list $\ell=1,2,3$ and the total over $\ell=1,\ldots,5$.  The Dirac columns list $k=1,2,3$ and the total over $k=1,\ldots,5$ with $N_D=18$.}
\label{tab:sv_partial}
\begingroup
\scriptsize
\setlength{\tabcolsep}{3pt}
\begin{ruledtabular}
\begin{tabular}{ccccccccc}
$a/M$ & $P_{\gamma,1}$ & $P_{\gamma,2}$ & $P_{\gamma,3}$ & $P_\gamma$ & $P_{D,1}$ & $P_{D,2}$ & $P_{D,3}$ & $P_D$ \\
0.00 & $3.2961\times10^{-5}$ & $6.6755\times10^{-7}$ & $1.0094\times10^{-8}$ & $3.3639\times10^{-5}$ & $7.0890\times10^{-4}$ & $2.7039\times10^{-5}$ & $5.3165\times10^{-7}$ & $7.3648\times10^{-4}$ \\
0.50 & $2.6349\times10^{-5}$ & $4.5747\times10^{-7}$ & $5.9227\times10^{-9}$ & $2.6812\times10^{-5}$ & $6.0169\times10^{-4}$ & $1.9839\times10^{-5}$ & $3.3506\times10^{-7}$ & $6.2187\times10^{-4}$ \\
1.00 & $1.1699\times10^{-5}$ & $1.1515\times10^{-7}$ & $8.4066\times10^{-10}$ & $1.1815\times10^{-5}$ & $3.3294\times10^{-4}$ & $6.4209\times10^{-6}$ & $6.1903\times10^{-8}$ & $3.3942\times10^{-4}$ \\
1.50 & $1.2783\times10^{-6}$ & $2.5489\times10^{-9}$ & $3.7002\times10^{-12}$ & $1.2809\times10^{-6}$ & $6.7554\times10^{-5}$ & $2.8883\times10^{-7}$ & $5.7541\times10^{-10}$ & $6.7844\times10^{-5}$ \\
1.90 & $6.1427\times10^{-10}$ & $9.7302\times10^{-15}$ & $1.3745\times10^{-19}$ & $6.1428\times10^{-10}$ & $2.8489\times10^{-7}$ & $1.0389\times10^{-11}$ & $2.0338\times10^{-16}$ & $2.8490\times10^{-7}$ \\
1.99 & $7.9078\times10^{-15}$ & $3.8642\times10^{-22}$ & $1.6831\times10^{-29}$ & $7.9078\times10^{-15}$ & $6.2811\times10^{-11}$ & $7.0418\times10^{-18}$ & $4.3357\times10^{-25}$ & $6.2811\times10^{-11}$ \\
\end{tabular}
\end{ruledtabular}
\endgroup
\end{table*}

\begin{table*}[t]
\centering
\caption{Direct CFM partial powers in the same notation as Table~\ref{tab:sv_partial}.}
\label{tab:cfm_partial}
\begingroup
\scriptsize
\setlength{\tabcolsep}{3pt}
\begin{ruledtabular}
\begin{tabular}{ccccccccc}
$a/M$ & $P_{\gamma,1}$ & $P_{\gamma,2}$ & $P_{\gamma,3}$ & $P_\gamma$ & $P_{D,1}$ & $P_{D,2}$ & $P_{D,3}$ & $P_D$ \\
0.00 & $1.4441\times10^{-3}$ & $3.0791\times10^{-4}$ & $5.0397\times10^{-5}$ & $1.8105\times10^{-3}$ & $0.0121$ & $4.2914\times10^{-3}$ & $8.7627\times10^{-4}$ & $0.01743$ \\
1.00 & $2.6774\times10^{-4}$ & $2.1484\times10^{-5}$ & $1.3072\times10^{-6}$ & $2.9061\times10^{-4}$ & $3.3634\times10^{-3}$ & $4.7205\times10^{-4}$ & $3.6343\times10^{-5}$ & $3.8741\times10^{-3}$ \\
1.50 & $3.2961\times10^{-5}$ & $6.6756\times10^{-7}$ & $1.0094\times10^{-8}$ & $3.3639\times10^{-5}$ & $7.0890\times10^{-4}$ & $2.7039\times10^{-5}$ & $5.3166\times10^{-7}$ & $7.3648\times10^{-4}$ \\
1.80 & $9.4806\times10^{-7}$ & $1.5055\times10^{-9}$ & $1.7219\times10^{-12}$ & $9.4957\times10^{-7}$ & $5.4312\times10^{-5}$ & $1.8830\times10^{-7}$ & $2.9864\times10^{-10}$ & $5.4500\times10^{-5}$ \\
1.90 & $3.7182\times10^{-8}$ & $6.5934\times10^{-12}$ & $9.2160\times10^{-16}$ & $3.7188\times10^{-8}$ & $5.4137\times10^{-6}$ & $2.2594\times10^{-9}$ & $4.4264\times10^{-13}$ & $5.4160\times10^{-6}$ \\
1.95 & $1.2419\times10^{-9}$ & $3.5542\times10^{-14}$ & $9.0206\times10^{-19}$ & $1.2419\times10^{-9}$ & $4.6517\times10^{-7}$ & $2.9771\times10^{-11}$ & $1.0343\times10^{-15}$ & $4.6520\times10^{-7}$ \\
1.97 & $1.1148\times10^{-10}$ & $1.0027\times10^{-15}$ & $8.0251\times10^{-21}$ & $1.1148\times10^{-10}$ & $7.8017\times10^{-8}$ & $1.5210\times10^{-12}$ & $1.6591\times10^{-17}$ & $7.8019\times10^{-8}$ \\
1.99 & $8.6878\times10^{-13}$ & $7.5089\times10^{-19}$ & $5.6718\times10^{-25}$ & $8.6878\times10^{-13}$ & $2.0446\times10^{-9}$ & $3.7737\times10^{-15}$ & $3.8743\times10^{-21}$ & $2.0446\times10^{-9}$ \\
\end{tabular}
\end{ruledtabular}
\endgroup
\end{table*}

\section{Adiabatic Half-Decay Estimate}

It is useful to ask whether an analogue of Table~II of Ref.~\cite{BolokhovKonoplya2025Circumvent} can be made with the present direct greybody factors.  Such an estimate has a limited but clear meaning.  It is not a complete back-reacting lifetime.  It assumes an adiabatic trajectory on the Simpson--Visser black-hole branch in which the dimensional parameter $a$ is held fixed while the mass decreases, so that
\begin{equation}
 \xi\equiv \frac{a}{M}
\end{equation}
monotonically grows.  With the instantaneous luminosity written as
\begin{equation}
 \frac{dM}{dt}=-\frac{\alpha(\xi)}{M^2},
 \qquad
 \alpha(\xi)=M^2\left(P_\gamma+P_D\right),
\end{equation}
the time needed to evolve from $\xi_i$ to $\xi_f$ is
\begin{equation}
\label{eq:direct_lifetime_integral}
 \tau=\frac{G^2a^3}{\hbar c^4}
 \int_{\xi_i}^{\xi_f}\frac{d\xi}{\xi^4\alpha(\xi)} .
\end{equation}
For direct comparison with Ref.~\cite{BolokhovKonoplya2025Circumvent}, we take $\xi_i=1$, $\xi_f=1.99$, $a=M_0$, and use the same right-endpoint quadrature over the published Simpson--Visser sample points $\xi=1.5,1.9,1.99$.  Replacing the WKB luminosity by the direct-GBF values in Table~\ref{tab:sv_power} gives
\begin{equation}
 \int_{1}^{1.99}\frac{d\xi}{\xi^4\alpha_{\rm dir}(\xi)}
 \simeq 9.14648\times10^7,
\end{equation}
and therefore
\begin{equation}
 \tau_{\rm dir}\simeq
 4.78339\times10^{-22}\left(\frac{M_0}{1\,{\rm g}}\right)^3 {\rm s} .
\end{equation}
The corresponding WKB coefficient obtained from Table~II of Ref.~\cite{BolokhovKonoplya2025Circumvent} is $5.60873\times10^{-24}{\rm s}$ for $M_0=1\,{\rm g}$.  The direct-GBF half-decay estimate is therefore about $85.3$ times longer, as shown in Table \ref{tab:direct_lifetime}.  The enhancement is driven almost entirely by the last interval near $\xi=1.99$, where the direct fermion luminosity is much smaller than the WKB value.

\begin{table}[t]
\centering
\caption{Adiabatic Simpson--Visser half-decay estimate from $a/M=1$ to $a/M=1.99$ at fixed dimensional $a=M_0$.  The WKB column reproduces the quadrature convention of Table~II in Ref.~\cite{BolokhovKonoplya2025Circumvent}; the direct column replaces $\alpha=P_\gamma+P_D$ by the numerical-GBF luminosities computed here.}
\label{tab:direct_lifetime}
\begingroup
\scriptsize
\begin{ruledtabular}
\begin{tabular}{cccc}
$M_0$ [g] & $\tau_{\rm WKB}$ [s] & $\tau_{\rm direct}$ [s] & $\tau_{\rm direct}/\tau_{\rm WKB}$ \\
$10^{0}$ & $5.6087\times10^{-24}$ & $4.7834\times10^{-22}$ & $85.28$ \\
$10^{5}$ & $5.6087\times10^{-9}$ & $4.7834\times10^{-7}$ & $85.28$ \\
$10^{10}$ & $5.6087\times10^{6}$ & $4.7834\times10^{8}$ & $85.28$ \\
$10^{15}$ & $5.6087\times10^{21}$ & $4.7834\times10^{23}$ & $85.28$ \\
\end{tabular}
\end{ruledtabular}
\endgroup
\end{table}

\section{Interpretation}

The direct calculation supports the central thermodynamic picture of Ref.~\cite{BolokhovKonoplya2025Circumvent}: as the black holes approach the macroscopic-wormhole endpoint, the Hawking temperature goes to zero and the instantaneous luminosity becomes very small.  The greybody thresholds do not move dramatically; instead, the thermal spectrum slides into the low-frequency tail of the same lowest barriers.  The residual particle flux is then dominated by fermions because the lowest Dirac barrier is lower and the Fermi--Dirac denominator is less suppressive at small frequency than the photon channel with its stronger low-frequency tunneling.

The new point is quantitative.  A WKB expansion is organized around the potential maximum and is most accurate for sufficiently high multipoles and frequencies near the barrier top \cite{IyerWill1987}.  Hawking luminosities near the endpoint are controlled by a different region: low $\omega$, low multipole number, and sometimes a long tortoise-coordinate tail.  This explains why the WKB method can reproduce the Schwarzschild row reasonably well while failing by orders of magnitude in the cold CFM endpoint.  The issue is not a disagreement about the temperature formula; it is a greybody-factor issue in the part of the spectrum that matters most for the integrated flux.

The CFM branch also illustrates that the sign of the WKB error is spin and parameter dependent.  At $a=0$, WKB overestimates the photon luminosity but underestimates the fermion luminosity.  Near $a=2M$, WKB overestimates both channels.  Therefore a single global correction factor cannot turn the WKB table into the direct table.  The direct scattering calculation is required if percent-level or better emission rates are needed.

\section{Limitations}

The present calculation treats electromagnetic and massless Dirac fields as test fields on fixed backgrounds.  It does not include gravitational perturbations or a back-reacting evaporation trajectory.  It also follows Ref.~\cite{BolokhovKonoplya2025Circumvent} by comparing instantaneous equilibrium powers at fixed $M=1$ and fixed $a/M$.  The half-decay estimate above is therefore only the specified fixed-$a$ adiabatic quadrature, not a unique physical lifetime.  A physical evaporation history would require specifying how $a$ changes as the mass decreases and how the geometry transitions through, or stops before, the limiting wormhole surface.

Mass thresholds are also ignored.  The Dirac sector is massless and uses the multiplicity normalization of the reference WKB calculation.  If the instantaneous temperature falls below a particle rest mass, that species should be removed or Boltzmann suppressed.  This caveat is especially relevant near the cold endpoint, where the spectra peak at very small $M\omega$.

Finally, the comparison with Ref.~\cite{BolokhovKonoplya2025Circumvent} uses their published integrated WKB luminosities rather than digitized WKB greybody curves.  This is sufficient for the main observable quantity, the emission rate, but it means that curve-by-curve WKB residuals are not reconstructed here.

\section{Conclusion}

We have computed direct electromagnetic and massless Dirac greybody factors for the Simpson--Visser and CFM black-hole families studied in Ref.~\cite{BolokhovKonoplya2025Circumvent}.  The calculation uses the same $M=1$ units, the same parameter values and the same fermion multiplicity normalization as the WKB analysis.  The Schwarzschild limit reproduces Page's luminosities, validating the direct-scattering normalization.

The qualitative result is robust: both families become cold as $a\to2M$, and the Hawking luminosity is strongly suppressed.  The residual massless flux becomes increasingly fermion dominated.  The direct greybody thresholds change smoothly and only moderately, so the suppression is mainly caused by the falling Hawking temperature rather than by a rapidly growing barrier.

The quantitative emission rates differ from the WKB values in important regimes.  For mild Simpson--Visser deformations the photon powers agree at the percent level and the fermion powers at the several-percent level.  For near-endpoint Simpson--Visser and especially for the CFM family, the differences become large.  In the cold CFM rows the published WKB luminosities overestimate the direct powers by factors ranging from several to more than a thousand, depending on spin and parameter.  In the fixed-$a$ Simpson--Visser half-decay estimate, this changes the lifetime coefficient by a factor of about $85$.  Thus direct greybody factors are essential for reliable evaporation rates in black holes that approach a wormhole endpoint.

It has recently become clear that the scattering properties of black-hole effective potentials and their quasinormal spectra are intimately related manifestations of the same underlying wave dynamics. In this context, the transmission and reflection amplitudes associated with a potential barrier encode nontrivial information about the corresponding resonant modes, establishing a connection between grey-body factors and quasinormal frequencies \cite{Konoplya:2024vuj,Konoplya:2024lir}. This relation suggests that the frequency dependence of grey-body factors can serve as an additional tool for investigating the damped oscillations responsible for the ringdown stage of perturbed black holes. Therefore, accurate scattering data may provide valuable insight into the quasinormal spectrum, supplementing conventional time-domain and frequency-domain approaches to black-hole perturbation theory \cite{Lutfuoglu:2025mqa,Dubinsky:2024vbn,Bolokhov:2024otn,Skvortsova:2024msa,Malik:2024cgb}.

The high-precision grey-body factors calculated in the present work offer an opportunity to test this correspondence. At the same time, it should be emphasized that the grey-body factor--quasinormal mode correspondence is not universal and may depend sensitively on the structure of the effective potential and the properties of the perturbing field \cite{Konoplya:2025ixm,Konoplya:2025hgp,Antoniou:2025bvg,Antoniou:2026jvh}. Consequently, its applicability must be examined individually for each gravitational background and perturbation sector.

Shortly before the submission of the present work to arXiv, a paper appeared in which grey-body factors were also computed numerically for several black-hole spacetimes, including the Simpson--Visser geometry \cite{Arbey:2026koc} (see also \cite{Calza:2026wuf}). Therefore, a certain degree of overlap between that study and our analysis is unavoidable. Nevertheless, the methodology employed in the two works is different, and a detailed comparison of the respective approaches and results would be of considerable interest. We leave such a comparative analysis for future investigation.

\section*{Declaration of Competing Interest}
The authors declare that they have no known competing financial interests or personal relationships that could have appeared to influence the work reported in this paper.

\section*{Data Availability}
No data was used for the research described in the article.

\begin{acknowledgments}
B. C. L. is grateful to the Excellence project FoS UHK 2205/2025-2026 for the financial support.
\end{acknowledgments}

\bibliographystyle{apsrev4-1}
\bibliography{HawkingCircumvent}

\end{document}